\documentclass{article}
\usepackage{aaai2026} 
\usepackage{times}  
\usepackage{helvet} 
\usepackage{courier} 
\usepackage[hyphens]{url} 
\usepackage{graphicx}
\usepackage{natbib} 
\usepackage{booktabs}
\usepackage{amsmath}
\usepackage[T1]{fontenc}
\usepackage{caption}
\usepackage{array}
\usepackage{algorithm}
\usepackage{algorithmic}

\usepackage{listings}
\usepackage{xcolor}
\lstdefinestyle{prompt}{
  basicstyle=\ttfamily\footnotesize,
  breaklines=true,
  breakatwhitespace=true,
  postbreak=\mbox{\textcolor{gray}{$\hookrightarrow$}\space},
  columns=fullflexible,
  keepspaces=true,
  showstringspaces=false,
  frame=single,
  framerule=0.4pt,
  rulecolor=\color{gray!50},
  framesep=4pt,
  xleftmargin=0pt,
  numbers=none,
  literate=
    {"}{{\textquotedbl}}1
}

\usepackage{newfloat}
\usepackage{listings}
\DeclareCaptionStyle{ruled}{labelfont=normalfont,labelsep=colon,strut=off} 
\floatstyle{ruled}
\newfloat{listing}{tb}{lst}{}
\floatname{listing}{Listing}
\title{Measuring and Mitigating Overreliance to Build Human-Compatible AI}
\author{
    Lujain Ibrahim\textsuperscript{\rm 1}\thanks{Corresponding author: lujain.ibrahim@oii.ox.ac.uk},
    Katherine M. Collins\textsuperscript{\rm 2},
    Sunnie S. Y. Kim\textsuperscript{\rm 3, 4},
    Anka Reuel\textsuperscript{\rm 5, 6},\\
    Max Lamparth\textsuperscript{\rm 5},
    Kevin Feng\textsuperscript{\rm 7},
    Lama Ahmad\textsuperscript{\rm 8},
    Prajna Soni\textsuperscript{\rm 9},
    Alia El Kattan\textsuperscript{\rm 10},
    Merlin Stein\textsuperscript{\rm 1, 11},\\
    Siddharth Swaroop\textsuperscript{\rm 6},
    Vishakh Padmakumar\textsuperscript{\rm 5},
    Ilia Sucholutsky\textsuperscript{\rm 10},
    Andrew Strait\textsuperscript{\rm 11},
    Diyi Yang\textsuperscript{\rm 5},\\
    Q. Vera Liao\textsuperscript{\rm 12},
    Umang Bhatt\textsuperscript{\rm 2}
}
\affiliations{
    \textsuperscript{\rm 1}University of Oxford\\
    \textsuperscript{\rm 2}University of Cambridge\\
    \textsuperscript{\rm 3}Princeton University\\
    \textsuperscript{\rm 4}Apple\\
    \textsuperscript{\rm 5}Stanford University\\
    \textsuperscript{\rm 6}Harvard University\\
    \textsuperscript{\rm 7}University of Washington\\
    \textsuperscript{\rm 8}OpenAI\\
    \textsuperscript{\rm 9}Alinia AI\\
    \textsuperscript{\rm 10}New York University\\
    \textsuperscript{\rm 11}UK AI Security Institute\\
    \textsuperscript{\rm 12}University of Michigan\\
}

\begin{document}

\maketitle

\begin{abstract}
Large language models (LLMs) distinguish themselves from previous technologies by functioning as collaborative ``thought partners,'' capable of engaging more fluidly in natural language on a range of tasks. As LLMs increasingly influence consequential decisions across diverse domains from healthcare to personal advice, the risk of overreliance---relying on LLMs beyond their capabilities---grows. This paper argues that measuring and mitigating overreliance must become central to LLM research and deployment. First, we consolidate risks from overreliance at both the individual and societal levels, including high-stakes errors, governance challenges, and cognitive deskilling. Then, we explore LLM characteristics, system design features, and user cognitive biases that together raise serious and unique concerns about overreliance on LLMs in practice. We also examine historical approaches for measuring overreliance, identifying three important gaps and proposing three promising directions to improve measurement. Finally, we propose mitigation strategies that can be pursued to ensure LLMs augment rather than undermine human capabilities.
\end{abstract}

\begin{figure*}
    \centering
    \includegraphics[width=\textwidth]{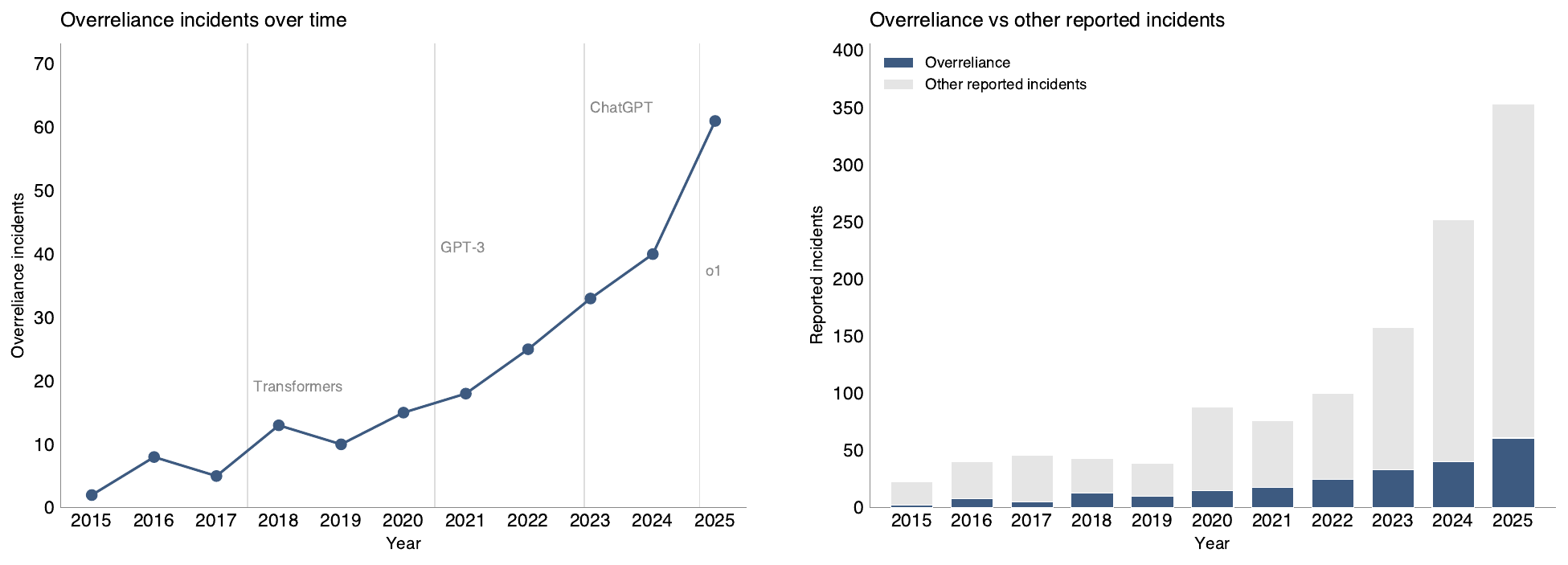}
\caption{\textbf{Reported incidents involving overreliance on AI systems, 2015--2025.} Incidents were extracted from the AI Incident Database ($N=1{,}339$) and labeled for whether they involved overreliance using unanimous agreement across three LLMs (GPT-4o-mini, GPT-4.1, Claude 4.5 Sonnet) with identical prompts. Overreliance was defined as cases where humans relied on AI systems (not exclusive to LLMs) without adequate verification, supervision, or critical evaluation. Vertical lines indicate major LLM capability milestones. On the left: absolute number of incidents classified as overreliance per year. On the right: overreliance incidents as a share of total reported incidents per year, with sample sizes ($n$) shown for each year. LLM judge ratings were validated by one of the study authors (see Appendix for system prompts and validation results).}

    \label{fig:figure1}
\end{figure*}

\section{Introduction}
Large language models (LLMs) are increasingly trusted with consequential decisions in domains such as healthcare, education, and personal life~\cite{ZaoSanders_2025, chatterji2025people}, where they support core cognitive functions like knowledge generation and reasoning~\cite{collins2024building}. However, our understanding of how humans interact with LLMs as thought partners lags behind their rapid adoption. This paper focuses on \textit{overreliance}---when users rely on LLMs beyond their warranted capabilities, accepting incorrect outputs or inappropriately delegating decisions~\citep{10.1145/3593013.3593978}. We argue that measuring and mitigating overreliance must become a central focus in LLM research and deployment. \looseness=-1

While the study of overreliance is well-established, the unique characteristics of LLMs---their conversational nature, complex text-based outputs, and versatility---complicate traditional approaches to measuring and mitigating overreliance and its impacts. For instance, LLM outputs can be partially correct or subjective. Their sycophantic behavior can lead users to misinterpret these outputs as definitive or accurate, particularly when they align with users' existing beliefs~\citep{sharma2023towards}. These issues may be exacerbated by automation bias and anthropomorphism, which are elevated to new risks by the fluency and potential persuasiveness of natural language outputs~\citep{abercrombie2023mirages}. \looseness=-1

More broadly, the general-purpose nature of LLMs can foster an illusion of universal competence~\citep{messeri2024artificial}, masking their domain-specific limitations and inconsistencies~\citep{zao-sanders2025}. This versatility also enables overreliance to extend beyond traditional task domains to emotional and social contexts, where users may overrely on LLMs for emotional support, relationship advice, or companionship~\citep{manzini2024code}. If overreliance is not measured, and steps are not taken towards its mitigation, LLMs, however capable, may fail to be human-compatible, preventing them from reaching their full potential in serving real users. They will be neither maximally helpful nor maximally harmless as many AI practitioners strive to build~\citep{huang2025position,sharma2026s}.

By systematically measuring and mitigating overreliance, we can better ensure LLMs augment human capabilities and enhance user autonomy, rather than diminish, displace, or homogenize them. To that end, this paper contributes the following: 
\begin{itemize}
    \item Grounding the study of overreliance on LLMs with an overview of the \textbf{history of overreliance research} across disciplines
    \item Articulating key \textbf{risks and potential negative impacts} of overreliance across individual and societal levels and near- and long-term horizons
    \item Identifying gaps in existing \textbf{measurement} approaches that arise from the unique characteristics of LLMs, and proposing new directions for measuring overreliance in human-LLM interactions
    \item Proposing promising directions for \textbf{mitigation} strategies for multiple research communities to address (technical AI research, policy, and human-computer interaction).
\end{itemize}

\section{A brief history of overreliance research}
Overreliance has been a subject of study across many fields, including human–computer interaction, cognitive science, psychology, computer science, and management science. It is commonly defined as either adopting a system output when that output is wrong, or delegating to a system when such delegation is undesirable~\citep{10.1145/3593013.3593978}. Specifically, overreliance is defined as a behavior, not a ``feeling nor an attitude, but the actual action conducted''~\citep{schemmer2023appropriate}. For example, an act of overreliance is committed when a programmer implements or accepts LLM-generated code suggestions that contain security vulnerabilities, as they defer to the system even if they have the information or expertise to act differently. 

\citet{parasuraman1997humans}'s typology of automation use categorizes disuse (underreliance) and misuse (overreliance) as opposite failures of human-machine interaction, as both forms can degrade interaction quality and outcomes. Underreliance could arise when users overestimate their own capabilities or distrust machines due to past errors, lack of interpretability, or pre-existing discomfort with automation~\cite{dietvorst2018overcoming,parasuraman2010complacency}. Dubbed ``algorithm aversion''~\citep{dietvorst2015algorithm}, individuals may ignore accurate recommendations or avoid using the system altogether, forfeiting potential benefits. Overreliance is frequently distinguished from trust, a related but distinct construct~\cite{dzindolet2003role,hou2021expert,zerilli2021how}; while trust represents a psychological state of willingness to be vulnerable based on positive expectations of others' actions~\cite{o2018linking}, overreliance manifests as the behavioral consequence of excessive trust~\citep{vereschak2021evaluate}. The relationship between trust and reliance is not linear~\cite{kahr2024trust,lee2004trust}. While trust guides the intention to rely, the action of reliance can also be influenced by other expectations and constraints such as saving time, reducing workload, avoiding risks, and exploring capabilities. 

Overreliance behaviors are often described as errors of commission, and they typically emerge without intent to offload responsibility, marking overreliance as an unintended outcome~\citep{passi2022overreliance}. To explain it, some researchers emphasize cognitive mechanisms behind overreliance, such as automation bias, complacency bias, and cognitive offloading~\citep{rastogi2022deciding}. Others focus on design or systemic factors. For instance, systems like current LLMs that are unable to reliably produce uncertainty estimates, frictionless interfaces, or organizations that treat AI systems as authoritative, can all foster overreliance by signaling that oversight is not possible or is unnecessary~\citep{cummings2006automation}. Despite differences in emphasis across disciplines, definitions converge on a few key points that we also center in this work: (1) overreliance is typically unintentional, (2) overreliance emerges from misaligned trust or expectations between the user and the AI system, and (3) overreliance produces measurable failures in judgment or oversight~\citep{schemmer2023appropriate}. \looseness=-1

\section{Risks from overreliance on LLMs}
\begin{table*}[t]
\centering
\caption{Short- and long-term impacts of AI overreliance across domains.}
\label{tab:overreliance-impacts}
\small
\begin{tabular}{p{2cm}p{6.5cm}p{6.5cm}}
\toprule
\textbf{Domain} & \textbf{Short-term} & \textbf{Long-term} \\
\midrule
Healthcare & 
\textbf{Individual:} Medical doctor follows incorrect AI diagnosis 
\par\medskip
\textbf{Institutional:} Hospital systems implement AI diagnostics without adequate verification protocols & 
\textbf{Individual:} Healthcare professionals experience cognitive deskilling and atrophy of diagnostic abilities 
\par\medskip
\textbf{Institutional:} Medical departments restructure workflows around AI systems, reducing human oversight \\
\addlinespace
\addlinespace
Personal advice & 
\textbf{Individual:} User accepts sycophantic or biased relationship/career advice without critical evaluation 
\par\medskip
\textbf{Societal:} Communities adopt similar AI-generated advice, creating homogenized decision patterns & 
\textbf{Individual:} Self-worth becomes derived from AI companion approval
\par\medskip
\textbf{Societal:} Social norms shift toward algorithmic validation of personal choices \\
\bottomrule
\end{tabular}
\end{table*}

Previous research on overreliance on technology has primarily focused on domains with clear evaluation criteria, where human and system outputs could be judged against established standards. In domains like medicine or programming, overreliance can be assessed against verifiable outcomes or technical correctness. However, when individuals defer to LLMs for life choices, creative decisions, or ethical judgments---areas without clear benchmarks for ``correct'' answers---overreliance becomes more difficult to detect yet potentially more consequential: uncritical agreement with LLMs can thus shape well-being, identity, and autonomy. As shown in Figure~\ref{fig:figure1}, reported AI incidents involving overreliance have increased with the broader AI deployment, and have overall represented a sustained share of undesirable AI incidents over the past decade. Here, we expand on these impacts of overreliance across domains and timescales.

\subsection{Individual risks}
\paragraph{Near-term: high-stakes errors}
In some instances, users may rely on AI systems for real-time information, or as inputs for decision-making, which can result in immediate errors or harm. This could occur in both personal and professional use. For example, an incorrect suggestion from an LLM used for a health-related decision could lead to a misdiagnosis or unsafe treatment plan, both in cases of self-diagnosis and misdiagnosis by a professional \cite{incidentdatabaseIncident838,choudhury2024large}. Similarly, unreliable information can be generated by LLMs in legal contexts, such as in a recent incident where lawyers submitted a court brief that unknowingly included six non-existent LLM-generated case citations \cite{mata2023avianca}. Furthermore, developers relying on AI-generated code without proper verification may end up deploying applications with critical security vulnerabilities. Such cases highlight that even skilled professionals may overrely on inaccurate outputs that sound plausible, especially when there is not adequate time or resourcing available to critically evaluate them \cite{nishal2024understanding}, or when there is limited education about where systems may be overly-confident or error prone \cite{long2020ai,vasconcelos2023explanations,ibrahim2023explanations}. As AI systems become more autonomous, self-improving, or deployed in open-ended environments, their brittleness and goal-alignment issues may lead to unprecedented failures.

\paragraph{Long-term: overreliance leading to dependency}
As users depend on AI systems to do their thinking, they risk losing both skills and motivation for independent problem-solving~\cite{gerlich2025human}, potentially creating a negative feedback loop of diminishing ability to calibrate appropriate trust~\cite{chan2023harms}. While overreliance describes instance-level failures of judgment, extended use of AI for tasks one could do oneself can produce overdependence, which is a pattern of substituting algorithmic processes for personal judgment, learning, and autonomy. The two can co-occur and reinforce each other, but each can also arise independently. Both can drive deskilling, a risk documented across domains including general problem-solving capacity \cite{zhai2024effects}, civic engagement \cite{kudina2024large}, and healthcare practice \cite{choudhury2024large}. Another critical domain is emotional overdependence on AI systems. Unlike task-based overreliance, this involves users developing inappropriate attachments to LLMs for emotional support, companionship, or relationship simulation, potentially displacing human connections \citep{ibrahim2026sycophantic, earp2025relational}. One can argue that narrow deskilling, such as losing mental arithmetic speed due to calculators, is less concerning because the tool is highly reliable, the skill is well-scoped, and the stakes of errors are typically low. LLMs, however, risk inducing \textit{broad} deskilling across interconnected cognitive capacities (e.g., critical reasoning, writing, judgment, and emotional processing). This is qualitatively different, first, because LLMs are not yet reliably correct across these domains, meaning the safety net that justifies narrow deskilling is absent. Second, these capacities are foundational.  Degradation in critical thinking, for instance, may affect not just LLM-assisted tasks, but one's ability to evaluate information and make decisions more generally.

\subsection{Societal risks}
\paragraph{Near term: governance challenges}     
At an institutional level, overreliance on LLMs can expose organizations to operational failures, decision-making biases, and cascading risks, particularly when human oversight is reduced or overlooked. For example, developers accepting flawed AI-generated code could lead to service disruptions, especially if there are no mechanisms in place to identify developers who fail to critically evaluate outputs~\cite{pu2025assistance,spatharioti2025effects}. Governance becomes especially challenging in critical sectors such as infrastructure, healthcare, and public safety, where multiple organizations may procure and deploy the same AI systems~\cite{balayn2024empirical}. Without sufficient oversight, shared vulnerabilities, overreliance, or overdependence patterns across institutions can amplify failures and cause widespread disruptions~\cite{cui2022ai}. Another example is in financial markets, where the reliance on LLM-generated news or stock analysis without proper regulatory oversight could lead to flash crashes if multiple systems react similarly to the same misinterpreted input~\cite{azzutti2024artificial}. The governance challenge here lies in ensuring that AI-driven systems comply with regulatory frameworks, especially in areas where biases or errors could have significant legal and reputational consequences~\cite{bhatt2024should}.

\paragraph{Long-term: influencing norms and homogenization}
Over time, AI use can generate new norms that shape behaviors, beliefs, and creative practices, essentially acting as new sources of ``culture'' or agents of social learning~\citep{brinkmann2023machine, collins2025revisiting}. One concern is that as organizations adopt LLMs for content creation (e.g., news articles, reports, policy drafts) or decision support, AI-generated outputs could unintentionally shape human discourse and norms. For instance, researchers have noted that overreliance on AI-generated content risks creating an ``echo chamber'' that stifles novel ideas and undermines the diversity of thought \cite{csernatoni2024}. This effect can scale when individual overreliance influences institutional-level outputs and decision-making. For example, randomized experiments show that LLM-assisted writers generate less diverse ideas and converge on similar phrasings, even after the AI is removed \cite{ashkinaze2024ai}. The combination of overreliance and overdependence on AI-generated content can flatten institutional and societal cultural practices, a phenomenon researchers have termed ``algorithmic monocultures'' or ``culture flattening''~\cite{chayka2025filterworld, kleinberg2021algorithmic}, which may become especially problematic as systems directly engage with us in new forms of ``collaborative cognition''~\citep{oktar2025identifyingevaluatingmitigatingrisks}.

\section{Factors influencing overreliance on LLMs}

To address these diverse risks, it is necessary to understand the factors that may underpin instances of overreliance. Previous work has proposed utility frameworks for modeling reliance decisions \citep{guo2024decision,wang2022will}, suggesting that reliance is influenced by factors affecting (1) perceived model correctness, (2) perceived self-correctness, and (3) evaluation of outcome payoffs (e.g., stakes and risk tolerance). We organize this section around model, system, and user factors to facilitate our later discussion of targeted mitigation strategies in Section~\ref{sec:mit}. \textit{Model characteristics} primarily influence perceived model correctness; \textit{system design} affects perceived model correctness and payoff evaluations; and \textit{user factors} shape perceived self-correctness and payoff evaluations. 

\begin{table*}[h!]
\centering
\caption{Selected factors influencing overreliance on LLMs.}
\label{tab:factors}
\begin{tabular}{ccccc}
\toprule
\parbox[c]{0.17\linewidth}{\centering\includegraphics[width=0.6cm]{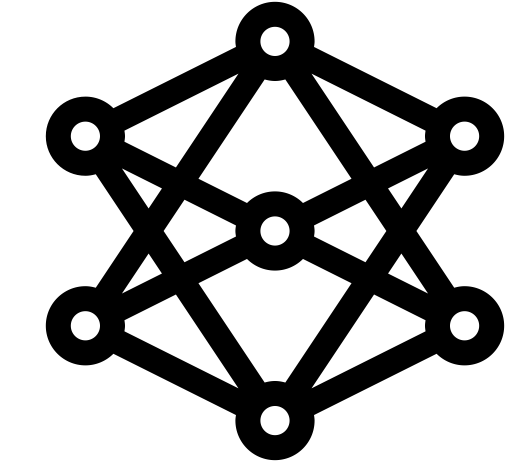} \textbf{Model}} & 
\parbox[c]{0.05\linewidth}{\centering$\boldsymbol{\leftrightarrow}$} &
\parbox[c]{0.17\linewidth}{\centering\includegraphics[width=0.6cm]{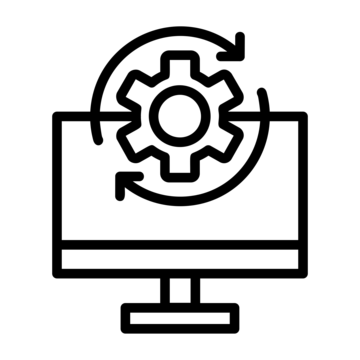} \textbf{System}} & 
\parbox[c]{0.05\linewidth}{\centering$\boldsymbol{\leftrightarrow}$} &
\parbox[c]{0.17\linewidth}{\centering\includegraphics[width=0.6cm]{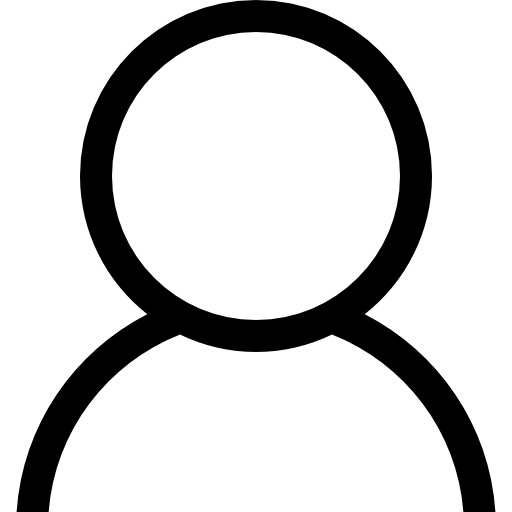} \textbf{User}} \\
\midrule
Anthropomorphic language & & Explanations & & Cognitive capacity \\
Sycophancy & & Interface design  & & User experience  \\
Certainty expressions & & Frictions & & Cognitive biases \\
\bottomrule
\end{tabular}
\end{table*}
\subsection{Model characteristics}
Several characteristics of LLMs can contribute to overreliance, most fundamentally through the anthropomorphic qualities of their natural language output. Language models simulate human-like conversational patterns, e.g., through asking follow-up questions or expressing helpfulness and empathy, influencing users' perceptions of their competence, trustworthiness, and alignment~\citep{abercrombie2023mirages, ibrahim2025multi, mckee2024warmth}. In some cases, users may develop an emotional attachment to models which could lead to increased reliance across both emotional and task-based domains, though this relationship remains an important area for future research \citep{manzini2024code}. Language models are also capable of expert mimicry, e.g., through citations and structured responses, assertive phrasing, or confident tone, which can further signal certainty and reinforce perceived authority \citep{metzger2024empowering}. Notably, language models can communicate hallucinations or fabricated content as factual statements, using the same fluent, human-like, and often confident language, making errors difficult to detect. Alongside social and expert mimicry, the tendency of models to be overly agreeable or sycophantic can reinforce users' existing beliefs and preferences, creating a feedback loop that discourages critical evaluation \citep{cheng2025social,sharma2023towards,ibrahim2025training}. Collectively, these characteristics can contribute to users conflating stylistic polish with epistemic reliability, as current models often default to confident, fluent outputs and are simultaneously unable to reliably express uncertainty that could prompt critical evaluation~\cite{zhou2024relying,yin2023large}. 

\subsection{System design}
Research in AI and human-computer interaction has investigated a wide range of interactive affordances that mediate human trust and reliance on AI \cite{amershi2019guidelines}. For example, the availability, or lack thereof, of (un)certainty indicators that communicate to users the likelihood of correctness of AI outputs has been shown to impact overreliance patterns \cite{kim2024m,ma2023should,zhou2024relying}. Notably, while emerging research has shown promise in token-level methods, sample-based methods \cite{shorinwa2024survey}, and multicalibration \cite{detommaso2024multicalibration}, novel challenges still remain in reliably computing (un)certainty of LLMs compared to their predictive predecessors. 
Another system affordance is \textit{adaptive explanations} which can tailor explanations of model outputs by task context or even other strategies such as levels of confidence \cite{lai2023selective, robrecht2023study, turchi2024adaptive}, e.g., explaining alternative predictions when AI confidence falls below certain thresholds, or only explaining top predictions when confidence is high \cite{ma2023should}. 
Interface designs, such as added friction or selective task delegation in multi-step agentic trajectories, have also been shown to influence reliance behaviors \cite{feng2024cocoa,he2025plan}.\looseness=-1

\subsection{User cognition and experience}
Several user factors may explain why a user overrelies on an LLM. Users with a reduced ``need for cognition''~\citep{cacioppo1982need} are more likely to overrely~\cite{buccinca2021trust}, while a user's disposition towards trusting and their general intuitions about the task at hand (e.g., their level of expertise) can influence baseline reliance behavior~\citep{chen2023understanding,hoff2015trust}. However, even without a predisposition to trust or with low confidence, users may still be swayed to overrely due to inherent cognitive limitations. Humans are constrained by time and cognitive capacity~\citep{GRIFFITHS2020873}, which is often addressed through ``resource-rational'' analyses~\citep{Lieder_Griffiths_2020}. These limitations suggest that when a task demands more than can be processed under our limited ``compute budget,'' users may default to relying on a model's output without proper verification~\citep{lee2025critical, sarkar2024intention,simkute2025ironies}. Propensity towards reliance is further shaped by user experience. Emerging research highlights a potential relationship between chatbot usage and reliance on model outputs, for example in social or affective contexts~\citep{fang2025ai, phang2025investigating}. While humans have the ability to update beliefs in light of new evidence~\citep{ griffiths2024bayesian,tenenbaum2011grow}, this updating process can go awry, with individuals giving undue weight to particularly salient experiences~\citep{chater2020probabilistic,tversky1974judgment}. In conversations, humans often make inferences about what is not said based on the structure of the conversation and come to these conversations with strong prior expectations about how they will unfold~\citep{Grice1975-GRILAC-6}. These expectations shape reliance decisions, with recent research showing that humans may struggle to generalize LLM successes and failures across domains, often miscalibrating their mental models of model behavior~\citep{ kelly2023capturing,steyvers2024three,vafa2024}.\looseness=-1

\section{Measuring overreliance on LLMs}
Long before ``reliance on AI'' emerged as a research field, reliance has been a subject of extensive measurement research across fields~\cite{lee1992trust,lee1994trust,lee2004trust,manzey2012human}. Much of this work investigates reliance on human advice using the \textit{judge-advisor framework}~\cite{bonaccio_dalal}, where a ``judge'' receives advice from an ``advisor'' and must decide what to do with the advice. Here, reliance is typically measured as ``advice acceptance,'' where studies have investigated various factors that influence this acceptance~\cite{Harvey1997,Sniezek2001,vanDongen}.

Recently, reliance on AI has been studied in AI-assisted decision-making scenarios following the same judge-advisor framework, where AI serves as the ``advisor'' and provides advice for the human ``judge.'' These studies, which surged around 2020, typically investigate discriminative models that predict a discrete label (e.g.,~\cite{buccinca2021trust,lai2021towards,zhang2020effect,yin2019understanding}). While a few recent studies investigate reliance on LLMs specifically, they focus on limited scenarios of making binary true/false judgments of factual statements~\cite{kim2024m,kim2025fostering,si2023large}. This kind of setup allows the treatment of reliance as a binary decision of accepting or rejecting the whole or a part of AI output in the final answer of the human ``judge''~\cite{hullman2025decision,lai2021towards}. \looseness=-1 

However, this conceptualization is an approximation that does not distinguish between situations where people's own judgment agrees with the AI output and ones where there is explicit reliance. 
Thus, other work instead measures reliance as ``switch fraction,'' which quantifies how often people change their initial answer to adopt the AI advice~\cite{He2023CHI,LuYin2019,yin2019understanding,zhang2020effect}, or ``weight of advice'' (WoA), which measures the degree to which people shift their answer toward the AI advice on a continuous scale~\cite{logg19people}. Amongst these measures, only WoA can be extended to measure reliance in a non-discrete context, such as in a numerical estimation task where one changes their own score to be closer to AI's score~\cite{bo2024rely}.

While the above-mentioned measures reflect how people descriptively rely on AI outputs, the measurement of \textit{appropriate} reliance is a more complex challenge. Appropriate reliance simultaneously requires relying on AI when the AI advice is desirable or helpful (\textit{not underrelying}, or correct AI reliance) and not relying on AI when the AI advice is undesirable or unhelpful (\textit{not overrelying}, or correct self-reliance). Past research often examines the two components separately by measuring reliance on desirable and undesirable AI outputs respectively~\cite{kim2024m,wang2021explanations}, or utilizes metrics that reflect the sum of the two~\cite{wang2021explanations}, the ratio of the two~\cite{bo2024rely}, or treats them as a two-dimensional metric~\cite{schemmer2023appropriate}. Some work bypasses quantifying overreliance or underreliance per se, but instead takes an outcome-oriented view by measuring, at an aggregate level, whether the human-AI joint outcome (e.g., joint decision accuracy) is better than that of the human alone (e.g.,~\cite{yin2019understanding,zhang2020effect}). 

\subsection{Why existing measures break down for LLMs}
As outlined above, traditional approaches to measuring reliance focus on experimental settings where humans interact with discriminative models that predict a discrete label. However, such approaches falter in the dynamic, conversational settings typical of human-LLM interactions. LLMs differ fundamentally from prior AI systems: their  \textbf{interactive use}, \textbf{complex text-based outputs}, and \textbf{general-purpose nature}, complicate how reliance, and thus overreliance, are measured. 

\paragraph{Interactive use}
LLMs undermine the foundational assumption of discrete decision points common in prior work. Rather than a single-shot interaction where a user accepts or rejects an AI advice, LLM usage involves iterative prompting and refinement, in which (over)reliance may unfold over time. As a result, paradigms like the judge-advisor framework~\cite{bonaccio_dalal,kim2024m,LuYin2019}, which require clean separations between user and AI contributions, struggle to capture the continuous and entangled nature of LLM use. Unlike earlier systems with traceable recommendations, LLMs operate through continuous back-and-forth exchanges, making it challenging to pinpoint the exact source of influence. Further, the influence of LLMs goes beyond direct suggestions, subtly shaping framing, language, and ideation over dialogue turns. 

\paragraph{Complex text-based outputs}

Traditional reliance metrics such as agreement~\cite{buccinca2021trust,He2023CHI,kim2024m,yin2019understanding} and switch fraction~\cite{He2023CHI,LuYin2019,yin2019understanding,zhang2020effect} presume binary correctness and a known ground truth---assumptions rarely met in open-ended generative tasks~\cite{passi2024appropriate}. Crucially, these metrics fail to capture the granular nature of reliance with LLMs, where the \textit{unit of reliance} can vary non-discretely across token, sentence, and topical levels. Users may rely on specific semantic elements while rejecting others within the same output, or trust the topical direction while questioning particular factual claims. This multi-layered reliance structure means LLM outputs may often be partially desirable, and users may engage with different components of them at varying levels of trust and critical evaluation. Additionally, contextual factors such as task type, user goals, domain expertise, and interaction quality influence how users engage with these complex LLM outputs \citep{chevi2025individual}. 

\paragraph{General-purpose nature}
LLMs are used across a wide spectrum of tasks, from information retrieval to creative writing, each with fundamentally different notions of what constitutes successful outcomes. This diversity challenges traditional concepts of ``appropriate reliance,'' as the definition of ``good'' performance shifts across contexts. In domains like programming or factual information retrieval, outputs can be evaluated against established standards of correctness. However, in areas like emotional support, relationship advice, or career guidance, success becomes subjective and context-dependent, with no clear benchmarks for ``correct'' answers~\citep{zao-sanders2025}. Without stable criteria for what constitutes good outcomes, determining appropriate levels of reliance becomes more challenging. Users may engage with LLMs in ways that affect their autonomy and well-being \citep{banker2019algorithm}, yet the absence of clear success metrics makes it difficult to evaluate when reliance becomes problematic. This blending of factual information and normative suggestions requires frameworks that acknowledge context-specific rather than universal notions of successful outcomes.

\subsection{Toward better measurement of overreliance on LLMs}
Given the shortcomings of existing measures in capturing reliance behaviors in user-LLM interactions, we outline three directions for developing better measures, each with concrete operationalizations. Table~\ref{tab:measurement} summarizes how LLM characteristics create measurement challenges and maps these to specific proposed metrics.

\begin{table*}[t]
\centering
\caption{Measurement framework: LLM characteristics, associated challenges, and proposed metrics.}
\label{tab:measurement}
\small
\begin{tabular}{p{2.8cm}p{3.5cm}p{4.2cm}p{4cm}}
\toprule
\textbf{LLM characteristic} & \textbf{Measurement challenge} & \textbf{Proposed metric(s)} & \textbf{Data source} \\
\midrule
Interactive, multi-turn use & No discrete decision points; reliance unfolds over time & Session-level adoption trajectory, turn-by-turn influence scores & Interaction logs, conversation transcripts \\
\addlinespace
Complex, partial outputs & Varying units of reliance (token, sentence, topic) & Edit distance, semantic similarity at multiple granularities, component-level acceptance & Pre/post output comparison, embeddings \\
\addlinespace
General-purpose nature & Context-dependent success criteria & Outcome-based assessment, domain-specific goal completion & Task completion data, delayed follow-up \\
\addlinespace
Subjective/open-ended tasks & No ground truth for ``correct'' reliance & User goal completion, well-being maintenance or enhancement, downstream behavioral indicators & Surveys, longitudinal tracking \\
\addlinespace
Persuasive, fluent outputs & Difficulty distinguishing quality from correctness & Calibration curves (reliance rate vs. actual correctness), verification behavior tracking & Controlled studies with planted errors \\
\bottomrule
\end{tabular}
\end{table*}

\paragraph{Go beyond single-point measures.} 
LLMs produce more complex outputs than traditional AI models. For example, it is common for an LLM output to be highly fluent and factually accurate but have gaps in logic. Take a simple decision-making task as an example: a user has to decide between option A and option B. While traditional AI models would recommend a single option (A or B), LLMs may provide supporting arguments for both options, refrain from providing a recommendation, or recommend a single option but with varying arguments and degrees of confidence. In this example, how should reliance be measured? If a user agrees with an LLM's argument for its recommendation but ultimately does not follow that recommendation, did the user rely on the LLM or not? What about vice versa? In other more open-ended and generation-based tasks (e.g., writing, coding, planning), single-point measures, such as agreement, may again fall short, as LLMs typically do not provide a single recommendation. For these tasks, measuring the extent to which a user adopts the LLM's output (e.g., the number of words, tokens, or suggestions) may be a more suitable measure of reliance~\cite{vasconcelos2023explanations}. More concretely, metrics like normalized edit distance between user drafts and final outputs, or semantic similarity scores computed at sentence and paragraph levels, can capture reliance at multiple granularities. For multi-turn interactions, tracking how reliance evolves across turns (including whether early LLM suggestions anchor later user decisions) may reveal patterns invisible to single-point measures. These methods can also follow other calls for richer ways of evaluating LLMs~\citep{ying2025benchmarking}. 

\paragraph{Shift from outputs to outcomes.}  In traditional reliance measurement, there is often a clear notion of ``good'' outputs, typically defined as correct classifications according to ground truth labels, based on which over-, under-, and appropriate reliance are defined.  In LLM use cases with verifiable outcomes like factual information retrieval, appropriate reliance can be assessed against a ground truth. However, LLMs are often used in subjective, open-ended domains where this may not be the case. Thus, we advocate for a shift in focus from the outputs of LLMs to the \textit{outcomes} that result from user interactions with these outputs (e.g., did the user achieve their goal? Was user well-being preserved?) when assessing the appropriateness of reliance. For a given task (e.g., getting correct information or producing a satisfactory piece of writing or code), ``appropriate reliance'' may be defined as relying on the LLM and achieving the desired outcome; ``overreliance'' as relying on the LLM and not achieving the desired outcome; and ``underreliance'' as not relying on the LLM and not achieving the desired outcome, when only relying on the LLM would have given the desired outcome. Importantly, this requires measuring outcomes at appropriate time horizons, i.e., immediately for factual tasks, but potentially days or weeks later for advice-seeking or planning tasks where consequences unfold over time.\looseness=-1

\paragraph{Employ additional measures to holistically understand reliance.} While adopting and adapting commonly used behavioral measures of reliance (e.g., agreement) and engaging with historical research on reliance will remain critical, employing additional measures is necessary to gain complementary in-depth insights to specific human-LLM interaction contexts. For example, when studying reliance on LLMs in information-seeking contexts, additional tracking of users' source clicking behavior and follow-up questions can provide deeper insights into when and how much users rely on LLMs and what factors shape their reliance~\cite{kim2024m,kim2025fostering}. Low verification behavior despite high reliance may signal overreliance even when traditional metrics suggest appropriate use. Other common additional measures include users' self-reported confidence in their final answer and evaluation of AI outputs: increased confidence may indicate increased reliance, and this may be miscalibrated \citep{vodrahalli2022uncalibrated}. Additionally, increased overreliance may lead to less learning and enjoyment of the task, which can also be measured~\citep{glikson2020human}. For affective use or personal advice applications of LLM, surveys and behavioral measures of user well-being, such as self-reported mood surveys, stress or anxiety scales, and behavioral indicators of decision satisfaction, can also be employed \citep{cao2022understanding}.

\section{Mitigating overreliance on LLMs}\label{sec:mit}
Once we can better measure overreliance, the next challenge is mitigation. Below, we outline several directions to that end.

\subsection{Model-level mitigations}
Attending to the linguistic and stylistic choices in language model training is important for managing overreliance at the model level. Current LLMs exhibit several patterns that may influence user perceptions of system capabilities and confidence. First, human-like language features, such as empathetic responses and social pleasantries, can lead users to attribute greater agency and competence to systems \cite{abercrombie2023mirages,ibrahim2025multi}. Second, LLMs often express high certainty: plain statements without epistemic markers are perceived as confident assertions, which may cause users to overestimate response reliability \cite{zhou2024relying}.
These patterns appear to emerge partly from training processes. Existing work suggests that RLHF can amplify sycophancy \citep{sharma2023towards}, reward modeling tends to favor certainty over accuracy, and human raters often prefer confident-sounding responses regardless of correctness \citep{zhou2024relying}. LLMs also appear to assume users are more rational than they typically are~\cite{liu2024large}.
Addressing these patterns may require targeted interventions. 
Improving models to accurately convey their uncertainty, in natural language (e.g., ``I'm not sure, maybe...'') or other forms (e.g., confidence scores), could help users calibrate reliance \cite{kim2024m,Vasconcelos_2024}.
Future research might explore modified reward functions that account for unwarranted certainty and human feedback protocols that encourage more balanced (un)certainty expressions. We also encourage that any modifications that are made to models are appropriately communicated if receiving feedback from users, e.g., with FeedbackLogs~\citep{barker2023feedbacklogs}, to ensure that users are aware when they may want to explicitly adjust their expectations for models.

\subsection{System-level mitigations}
Interface design choices that incorporate cognitive forcing functions and visual cues can help mitigate overreliance. For example, priming statements that clarify the model’s goals and limitations can set expectations of model capabilities \cite{collins2024modulating,pataranutaporn2023influencing}. Mixed-initiative controls, where both the user and AI suggest, produce, evaluate, and modify outputs in response to each other, can help establish feedback loops that promote balanced interaction \cite{ deterding2017mixed, feng2024cocoa, 10.1145/302979.303030, subramonyam2024bridging}. AI-assisted autocomplete can allow users to phrase queries more effectively, bridging the gap between user intent and system capabilities \cite{subramonyam2024bridging}. Friction mechanisms, like an extra click or disclaimers, can also slow users down at critical decision points, further reducing reliance on AI outputs \cite{collins2024modulating,hunter2024monitoring,naiseh2021nudging,weisz2024design}. Finally, visual cues like typing indicators or word-by-word output, which contribute to the illusion of human-like behavior, can be revised or minimized depending on the use case \cite{abercrombie2023mirages}. These system-level interventions, rather than modifications to the model itself, may be able to effectively foster appropriate user reliance. 

\subsection{User-level mitigations}
Educating users about AI systems, how they work, their capabilities, and their limitations, can empower individuals to make more informed decisions about when to rely on AI outputs~\citep{chen2025chiea,long2020ai}. Such interventions should ideally cater to users’ existing knowledge, be presented in an engaging manner, and be broadly accessible. For example, recent AI literacy efforts have utilized games as medium to communicate to younger users~\citep{csapo2019survival}. 
And, recent studies have found users often overrely on LLMs and are unaware of how their outputs could be flawed (e.g., LLMs can provide unfaithful and inconsistent explanations)~\citep{kim2024m,kim2025fostering}; correcting such misconceptions could be engineered as part of AI literacy efforts.
Further, while substantial effort has gone into benchmarking AI systems’ performance on various tasks, these benchmarks may not be immediately interpretable to the general public (e.g., what does an MMLU accuracy of 85\% mean to an everyday user?) and warrant simplified summaries around expected performance. We may imagine, for instance, AI literacy efforts could contribute reports detailing expected model behavior to help correct for errors in our ``generalization functions''~\citep{vafa2024}.\looseness=-1

\section{Discussion}

We close by addressing three tensions that arise in discussions of overreliance on LLMs, each of which informs why we view measurement  as a central and urgent research priority.

\subsection{The current lack of empirical evidence}
A reasonable starting concern is that overreliance on LLMs may not be a pressing issue in practice, given the general lack of empirical evidence. In other words, existing instances of harm from inaccurate LLM outputs or emerging dependencies on LLM-based systems are often isolated cases. However, we argue that limited empirical evidence at this time is not indicative of \textit{no} issue of overreliance. Rather, it can be attributed, in part, to the absence of valid and appropriate methods for measuring overreliance in user-LLM interactions—as the saying goes, ``what is not measured cannot be improved''~\cite{kim2024m,kim2025fostering}. When deployment occurs outside of controlled experimental settings, we cannot afford to wait for problematic or prevalent cases to emerge as evidence before taking action, especially for long-term harms such as deskilling, the erosion of critical thinking abilities, or emotional dependence~\cite{casper2025pitfalls}. LLMs in their popular ``chatbot'' form remain early in their technological lifespan (circa 2022), and the identified instances of harms, even if currently limited, should be viewed as important signals to intervene and catalyze work on measuring and mitigating potential risks~\citep{Boyd_2025}.

\subsection{The role of advancing AI capabilities}
A second tension concerns whether overreliance is a transient problem. One might expect that as AI systems become increasingly sophisticated, overreliance concerns will naturally diminish: with hypothetically ``superhuman'' AI that rarely makes mistakes, traditional cases of overreliance would become increasingly rare. Paradoxically, however, as AI becomes more capable, detecting when these systems err becomes more difficult, as their outputs appear increasingly convincing~\cite{bender2021dangers}. Even as capabilities accelerate, these systems will still require human oversight for legal and societal reasons~\citep{bhatt2025learning,steyvers2022bayesian}. Advancing AI capabilities thus make measuring overreliance \textit{more} essential, not \textit{less}, to deploying AI for the benefit of all.

\subsection{Societal vs.\ technical interventions}
A final tension concerns where interventions should be located. One view holds that addressing overreliance through technical improvements or human-AI interaction design will ultimately prove fruitless, because the core challenges exist at societal and institutional levels. We agree that societal resilience and regulation are a critical piece of the puzzle, but contend that they hit a ceiling without complementary technical advances in measurement and system design. Regulations, for instance, cannot be effectively implemented without precise metrics to determine when and where overreliance is occurring. Educational and user empowerment initiatives may similarly face diminishing returns when systems themselves are designed in ways that inherently encourage overreliance~\citep{lu2024awareness}. A comprehensive approach therefore requires technical advances in measuring and mitigating overreliance.

\section{Conclusion}
LLMs distinguish themselves from previous technologies in their ability to engage in open-ended cognitive processes, functioning less as tools and more as collaborative thought partners~\citep{collins2024building, ibrahim2024beyond}. In this paper, we chart a path to address \textit{overreliance} on LLMs, showing that its consequences range from immediate errors to gradual shifts in human autonomy, skill development, and well-being. We call on the AI community to treat overreliance as a central research priority: developing new measures that can capture reliance across interaction turns and output granularities, building systems that convey calibrated uncertainty and introduce friction at critical decision points, and establishing deployment practices that preserve user autonomy. As LLMs seep into all aspects of our daily lives, proactively addressing overreliance will ensure these powerful technologies augment our capabilities instead of inadvertently---or worse, irreversibly---diminishing them.

\bibliography{aaai2026}

\appendix

\section{LLM-as-judge prompts}\label{app:prompts}

Each incident's title and description from the AI Incident Database
($N=1{,}339$) were classified independently by three large language
models: GPT-4o-mini, GPT-4.1, and Claude Sonnet~4.5
(\texttt{claude-sonnet-4-5-20250929}). All three models received the
\emph{identical} system prompt and user message template reproduced below,
with \texttt{temperature}~$=0$ and \texttt{max\_tokens}~$=50$. An incident
was labeled as involving overreliance only if \emph{all three} models
returned \texttt{\{"overreliance": true\}} (unanimous agreement); any
disagreement was treated as a negative label. Of the 1{,}339 incidents,
the unanimous rule labeled 147 as overreliance.

\medskip
\textbf{System prompt}
\medskip
\begin{lstlisting}[style=prompt]
You are an expert analyst identifying AI 
incidents that involve "overreliance on AI."
Overreliance on AI means situations where:
- Humans trusted AI systems too much 
without adequate verification or oversight
- Users blindly followed AI recommendations
without critical evaluation
- Lack of human supervision led to harm 
when AI made errors
- People deferred to AI judgment when they 
should have intervened
- Automation complacency - humans became 
inattentive because they trusted 
the automation
- Users did not question or verify AI 
outputs that turned out to be wrong
- AI was used as sole decision-maker when 
human judgment was needed


This does NOT include:
- AI systems that simply made errors 
(without human overreliance component)
- Algorithmic bias issues (unless humans 
failed to verify biased outputs)
- Privacy violations
- AI misuse/malicious use (scams, fraud, 
deepfakes used to deceive victims)
- Deepfake scams where victims were tricked 
by AI-generated content they didn't know 
was AI
- Technical failures without human 
oversight element
- Cases where someone was deceived BY AI 
(e.g. voice cloning
  scams) rather than overrelying ON AI
Respond with ONLY a JSON object:
{"overreliance": true} or {"overreliance": 
false}
\end{lstlisting}
\medskip
\textbf{User message}
\medskip
\begin{lstlisting}[style=prompt]
Analyze this AI incident for overreliance:

Title: {title}
Description: {description}

Is this an example of overreliance on AI?
\end{lstlisting}

\subsection{Validation}
To assess the reliability of the LLM-derived labels, we drew a stratified
random sample of 100 incidents (50 unanimously labeled as overreliance,
50 not), shuffled, and hand-coded each by one author who was blind to
the LLM votes. 
\begin{table}[h]
\centering
\begin{tabular}{lr}
\toprule
Sample size & 100 \\
Accuracy & 0.780 \\
Cohen's $\kappa$ & 0.528 \\
Precision & 0.703 \\
Recall & 0.703 \\
F1 & 0.703 \\
\midrule
Human positive rate & 37.0\% \\
LLM ensemble positive rate & 37.0\% \\
\bottomrule
\end{tabular}
\caption{Agreement between the unanimous LLM label and a hand-coded
human label on a stratified random sample of 100 incidents (50 drawn
from the consensus-positive class and 50 from the consensus-negative
class). $\kappa$ corresponds to ``moderate'' agreement.}
\end{table}
This moderate level of agreement supports the use of these LLM judge labels for a descriptive analysis of trends, similar to what we report here. Stronger quantitative claims (e.g., comparing rates across categories or testing for significant change at a specific time points) would benefit from additional validation, including using multiple human raters and a larger validation sample.
\end{document}